\documentclass[aps,showpacs,pra,amssymb, amsmath, twocolumn]{revtex4}
\usepackage{graphicx}
\usepackage{amssymb}
\usepackage{amsmath}
\usepackage{bm}
\usepackage{xcolor} 
\usepackage{array}
\usepackage{multirow}
\usepackage{tabularx}
\usepackage{booktabs}
\usepackage{textcomp}

\newcommand{\bra}[1]{\ensuremath{\langle #1 |}}
\newcommand{\ket}[1]{\ensuremath{| #1 \rangle}}

\begin{document}
\title{Quantum-Ising Hamiltonian programming in trio, quartet, and sextet qubit  systems}

\author{Minhyuk Kim$^1$, Yunheung Song$^1$, Jaewan Kim$^2$, and Jaewook Ahn$^1$}
\address{$^1$Department of Physics, KAIST, Daejeon 34141, Korea\\ 
$^2$Department of Physics, Myongji University, Yongin 17058, Korea}

\begin{abstract} Rydberg-atom quantum simulators are of keen interest because of their possibilities towards high-dimensional qubit architectures. Here we report three-dimensional conformation spectra of quantum-Ising Hamiltonian systems with programmed qubit connections. With a Rydberg-atom quantum simulator, various connected graphs, in which vertices and edges represent atoms and blockaded couplings, respectively, are constructed in two or three-dimensional space and their eigenenergies are probed during their topological transformations. Star, complete, cyclic, and diamond graphs, and their geometric intermediates, are tested for four atoms and antiprism structures for six atoms. Spectroscopic resolution ($\Delta E/E$) less than 10\% is achieved and the observed energy level shifts and merges through structural transformations are in good agreement with the model few-body quantum-Ising Hamiltonian.
 \end{abstract}

\maketitle

\section{Introduction}
Well-calibrated quantum many-body systems are currently in high demand because of their essential necessities for quantum applications such as quantum computing and quantum simulation~\cite{Feynman1982,Nielsen2000,Georgescu2014}. Among many promising physical platforms~\cite{Duan2010, Wendin2017, briegel2000,Gross2017, Weiss2017}, Rydberg-atom quantum simulators, which use a mesoscopic-scale, deterministic arrangement of neutral atoms with controllable strong local interactions induced by Rydberg-atom excitation, draw latest attentions~\cite{Saffman2016}. They have well-defined energy levels, relatively long coherence and lifetimes, and entanglements in these systems are generated with relative ease through giant dipole-dipole couplings in the Rydberg-atom blockade regime~\cite{Lukin2001,wilk2010,isenhower2010}. In recent demonstrations, these systems are used to generate as-many-as 20 qubit GHZ entangled states~\cite{Omran2019}, to observe the quantum many-body phenomena involved with localization~\cite{Marcuzzi2017} and thermalization~\cite{kim2018}, and also to investigate the critical phenomena of Ising-type or $XY$ quantum spin models across phase transitions~\cite{Bernien2016,Sylvain2019}. Of particular importance in the context relevant to the present work, Rydberg-atom quantum simulators are expected to realize the possibilities of high-dimensional qubit architectures~\cite{Saffman2016, kim2016,barredo2016,endres2016,Browaeys2020}.

Quantum simulation uses a controllable Hamiltonian $H$ of a quantum $N$-body system to reproduce or predict the behavior of other equivalent systems of model Hamiltonian $H_G$. The system size $N$ increases computational power exponentially, and the controllability in $H$ increases the simulation accuracy as well as the diversity of the systems to be simulated. Currently, Rydberg-atom quantum simulators have scaled up the number of qubits~\cite{Bernien2016}, approaching the  regime of quantum supremacy~\cite{Arute2019}. In that regards, dexterous control of the parameters of $H$, as many as possible and with high fidelities, is of great importance for a programmable quantum simulator. In this paper, we use a Rydberg-atom quantum simulator, of Hamiltonian $H$, to produce and tune a set of two- or three-dimensional arrangements of atoms, that are isomorphic to connected-graphs $\{G\}$ of $N=3-6$ qubits, to model quantum-Ising Hamiltonian $H_G$, and probe their topology-dependent eigenspectra, in particular, during structural transformations from one $G$ to another.

\section{Quantum-Ising Hamiltonian}
\label{theory1}

The Hamiltonian of $N$ atoms that are coherently excited to a Rydberg energy state is given by
\begin{equation}
\label{H1}
\hat{H}=\frac{\hbar\Omega}{2}\sum_{j=1}^N ( \ket{1}_j\bra{0}_j +  \ket{0}_j\bra{1}_j ) + \sum_{j<k} U({r}_{jk}) \hat{n}_j \hat{n}_k,
\end{equation}
where $\ket{0}_j$ and $\ket{1}_j$ denote the ground and Rydberg energy states, respectively, of an atom $j$ located at $\vec{r}_j$, $\Omega$ is the Rabi oscillation frequency, $U(r_{jk})=C_6/|\vec{r}_j-\vec{r}_k|^6$ is the van der Waals interaction between two Rydberg atoms, and $\hat{n}_{j} =\ket{1}_j\bra{1}_j$ is the excitation number~\cite{Lukin2001}. In the following, we will consider arrangements of atoms (e.g., see Fig.~1), in which only the nearest neighboring pairs, of the same inter-atom distance $d$, are within the Rydberg-blockade radius, i.e., $d<r_b=|C_6/\hbar\Omega|^{1/6}$, and thus all other long-ranged pairs are ignorable. Such an arrangement of atoms can be isomorphically represented by an undirected connected graph $G(V=N,E)$, in which the vertices $V$ denote atoms and the edges $E$ the nearest-neighbor couplings. Then, for a graph $G$, the Hamiltonian $H$ is given by the quantum-Ising Hamiltonian $H_G$ (with inhomogeneous longitudinal field):
\begin{equation}
\label{H2}
\hat{H}_G=J \sum_{(j,k) \in E} \hat{\sigma}_z^{(j)} \hat{\sigma}_z^{(k)} 
+ \sum_j^N \left( h_x \hat{\sigma}_x^{(j)} + h_z^{(j)} \hat{\sigma}_z^{(j)}\right),
\end{equation}
in which $\hat{\sigma}_{x,z}$ are Pauli spins, $J=U(d)/4$ is the coupling,  $h_x=\hbar\Omega/2$ is the transverse field, and $h_z^{(j)}=-||E_j||U(d)/2$  is the longitudinal field, with $||E_j||$ the number of edges for vertex $j$. 
So, the Hamiltonian $H$ of the Rydberg-atom quantum simulator is adequately modeled by the quantum-Ising Hamiltonian $H_G$ for a connected graph $G(V,E)$ that isomorphically represents an atom arrangement of equal nearest-neighbor couplings and ignorable long-range interactions. 
 
In order to investigate the topological change of a strongly-coupled Rydberg-atom system, we consider, for example, four-atom ($N=4$) arrangements and their structural transformations. As shown in Fig.~1(a), there are four nonisomorphic 4-vertex-connected graphs: the star graph, denoted by $S_4$, has one atom at the center and three at the ends of three claws, and three nearest-neighbor edges ($E=3$); the complete graph, $K_{4}$, has four atoms in the tetrahedron configuration with six edges ($E=6$); the cyclic graph, $C_4$, is the square configuration with four edges ($E=4$), and the diamond graph, $K_4$-e, has five edges ($E=5$). Here, $S_4$, $C_4$, and $K_{4}$-e are two-dimensional (e.g., in the $xy$ plane) and $K_{4}$ is three-dimensional, of the tetrahedron shape. Structural transformations among them can be proceeded, for example, in the sequence of $S_4\rightarrow K_4$, $K_4\rightarrow C_4$, and $C_4\rightarrow K_4$-e, as respectively shown in Figs.~1(b-d). 

\begin{figure}[tb]
\centering
\includegraphics[width=0.4\textwidth]{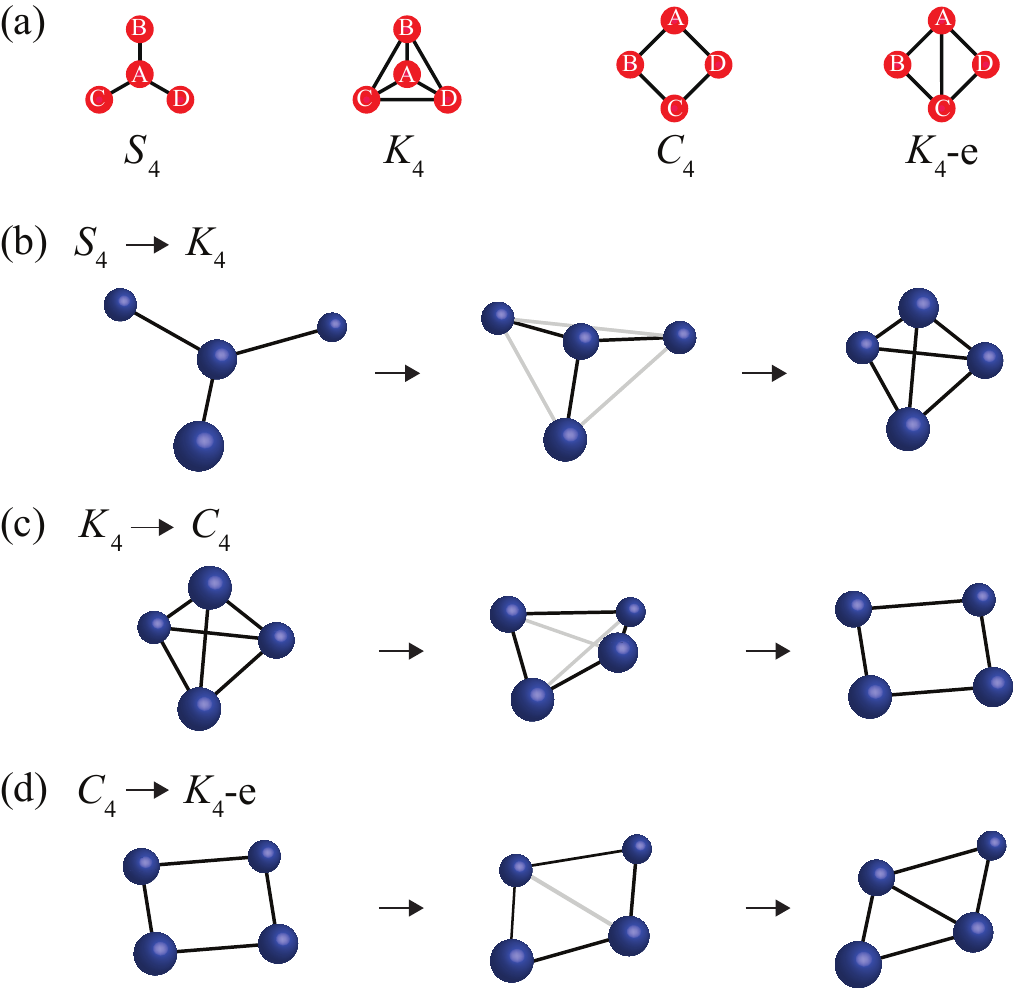}
\caption{(Color online) (a) Connected graphs for $N=4$ atom arrangements: $S_4$, the star graph, $K_4$, the complete graph, $C_4$, the cycle graph, and $K_4$-e, the diamond graph, in which vertices and edges represent atoms and Rydberg blockade couplings, respectively. (b-d) Structural changes (b) from $S_4$ to $K_4$, (c) from $K_4$ to $C_4$, (d) from $C_4$ to $K_4$-e.}
\label{Fig1}
\end{figure}

As a pedagogical example, let us consider $C_4$, the cyclic graph, which has four nearest-neighbor couplings. Due to the symmetry of the given graph, most of the possible $2^N=16$ eigenstates of $H_G$ are dark states, inaccessible from the initial state, $\ket{\Psi(t=0)}=\ket{0000}$, the bare-atom ground state, in our consideration. There are three bright eigenstates, of respective energies (in $\hbar=1$ unit hereafter) $\lambda_1=-\sqrt{\frac{3}{2}}\Omega$, $\lambda_5=0$, and $\lambda_7=\sqrt{\frac{3}{2}}\Omega$, where the index $j$ denotes the energy ordering among all energy states, bright and dark. The corresponding eigenstates are given by
\begin{subequations}
\label{allequations}
\begin{eqnarray}
\ket{\lambda_1} &=& -\frac{1}{\sqrt{3}}\ket{W_0}+\frac{1}{\sqrt{2}}\ket{W_1}-\frac{1}{\sqrt{6}}\ket{W_2^{C}} \\
\ket{\lambda_5} &=& -\frac{1}{\sqrt{3}}\ket{W_0}+\sqrt{\frac{2}{3}}\ket{W_2^C} \\
\ket{\lambda_7} &=& \frac{1}{\sqrt{3}}\ket{W_0}+\frac{1}{\sqrt{2}}\ket{W_1}+\frac{1}{\sqrt{6}}\ket{W_2^{C}},
\end{eqnarray}
\end{subequations}
which are represented with symmetric base states, $\ket{W_0}=\ket{0000}$, $\ket{W_1}=(\ket{1000}+\ket{0100}+\ket{0010}+\ket{0001})/\sqrt{4}$, and $\ket{W_2^{C}}=(\ket{1010}+\ket{0101})/\sqrt{2}$, each labeled with an excitation number (the number of atoms in the Rydberg-state). Likewise, there are four bright eigenstates for $S_4$, two for $K_4$,  and four for $K_4$-e. Table~I summarizes the eigenenergies and eigenstates of $H_G$, for the all four graphs.

\begin{table*}[t]
\centering
\caption{Quantum-Ising eigenstates of $N=4$ atom systems (bright states only), in symmetric base states defined by $\ket{W_0}=\ket{0000}$, $\ket{W_1} = (\ket{1000}+\ket{0100}+\ket{0010}+\ket{0001})/\sqrt{4}$, $\ket{W_2^C}=(\ket{1010}+\ket{0101})/\sqrt{2}$, $\ket{W_1^S} = (\ket{0100}+\ket{0010}+\ket{0001})/\sqrt{3}$, $\ket{W_2^S}=(\ket{0110}+\ket{0011}+\ket{0101})/\sqrt{3}$, $\ket{W_1^D} =  (\ket{1000}+\ket{0010})/\sqrt{2}$, and $\ket{W_1'^D} =  (\ket{0100}+\ket{0001})/\sqrt{2}$.}
\label{table1}
\begin{ruledtabular}
\begin{tabular}{lll}
Configuration & Eigenenergies ($\hbar=1$) & Eigenstates (bright states only) \\
\hline
Star  graph $S_3$ & 
$\lambda_1=-\sqrt{\frac{23}{10}}\Omega$ & 
$\ket{\lambda_1}= \sqrt{\frac{3}{20}} \ket{W_0}  -\sqrt{\frac{11}{30}} \ket{W_1^S} -\sqrt{\frac{1}{60}}\ket{1000} +\sqrt{\frac{7}{20}}\ket{W_2^S}  -\sqrt{\frac{7}{60}}\ket{0111} $ \\
& $\lambda_2=-\sqrt{\frac{10}{23}}\Omega$ & 
$\ket{\lambda_2}= -\sqrt{\frac{7}{20}}\ket{W_0}+\sqrt{\frac{1}{30}}\ket{W_1^S}+\sqrt{\frac{1}{5}}\ket{1000}+\sqrt{\frac{3}{20}}\ket{W_2^S}-\sqrt{\frac{4}{15}}\ket{0111}$ \\
& $\lambda_8=\sqrt{\frac{10}{23}}\Omega$ & 
$\ket{\lambda_8}= \sqrt{\frac{7}{20}} \ket{W_0}+\sqrt{\frac{1}{30}}\ket{W_1^S}+\sqrt{\frac{1}{5}}\ket{1000}-\sqrt{\frac{3}{20}}\ket{W_2^S}-\sqrt{\frac{4}{15}}\ket{0111}$ \\
& $\lambda_9=\sqrt{\frac{23}{10}}\Omega$ & 
$\ket{\lambda_9}= \sqrt{\frac{3}{20}}\ket{W_0}+\sqrt{\frac{11}{30}}\ket{W_1^S}+\sqrt{\frac{1}{60}}\ket{1000}+\sqrt{\frac{7}{20}}\ket{W_2^S}+\sqrt{\frac{7}{60}}\ket{0111}$ \\
\hline
Complete  graph $K_4$ & 
$\lambda_1=-\Omega$ & 
$\ket{\lambda_1}= \sqrt{\frac{1}{2}}\ket{W_0}-\sqrt{\frac{1}{2}}\ket{W_1}$ \\
& $\lambda_5=\Omega$ & 
$\ket{\lambda_5}= \sqrt{\frac{1}{2}}\ket{W_0}+\sqrt{\frac{1}{2}}\ket{W_1}$ \\
\hline
Cyclic graph $C_4$ & 
 $\lambda_1=-\sqrt{\frac{3}{2}}\Omega$ &
$\ket{\lambda_1}= -\sqrt{\frac{1}{3}}\ket{W_0}+\sqrt{\frac{1}{2}}\ket{W_1}  -\sqrt{\frac{1}{6}} \ket{W_2^C}$ \\
& $\lambda_5=0$ & 
$\ket{\lambda_5}= -\sqrt{\frac{1}{3}} \ket{W_0}+\sqrt{\frac{2}{3}} \ket{W_2^C}$ \\
&  $\lambda_7=\sqrt{\frac{3}{2}}\Omega$  & 
$\ket{\lambda_7}= \sqrt{\frac{1}{3}}\ket{W_0}+\sqrt{\frac{1}{2}}\ket{W_1}+\sqrt{\frac{1}{6}}\ket{W_2^C}$ \\
\hline
Diamond  graph $K_4$-e & 
$\lambda_1=-\sqrt{\frac{13}{10}}\Omega$ & $\ket{\lambda_1}= -\frac{3}{5}\ket{W_0}+\frac{3}{5}\ket{W_1^D} +\sqrt{\frac{7}{50}}\ket{W_1'^D} -\sqrt{\frac{7}{50}}\ket{1010}$ \\
& $\lambda_2=-\sqrt{\frac{5}{26}}\Omega$ & 
$\ket{\lambda_2}= -\sqrt{\frac{7}{50}}\ket{W_0}-\sqrt{\frac{7}{50}}\ket{W_1^D}+\frac{3}{5}\ket{W_1'^D}+\frac{3}{5}\ket{1010}$ \\
& $\lambda_{5}=\sqrt{\frac{5}{26}}\Omega$ & 
$\ket{\lambda_{5}}= \sqrt{\frac{7}{50}}\ket{W_0}-\sqrt{\frac{7}{50}}\ket{W_1^D}+\frac{3}{5}\ket{W_1'^D}-\frac{3}{5}\ket{1010}$ \\
& $\lambda_6=\sqrt{\frac{23}{10}}\Omega$ & 
$\ket{\lambda_6}= \frac{3}{5}\ket{W_0}+\frac{3}{5}\ket{W_1^D}+\sqrt{\frac{7}{50}}\ket{W_1'^D}+\sqrt{\frac{7}{50}}\ket{1010}$ \\
\end{tabular}
\end{ruledtabular}
\end{table*}

\section{Rydberg-atom quantum simulator}
In order to probe the eigenspectra of an $N$-body quantum-Ising Hamiltonian, we used a Rydberg-atom quantum simulator, which can (i) arrange $N$ single atoms isomorphically to a connected graph, (ii) implement the Hamiltonian $H$ in Eq.~\eqref{H1} through creating Rydberg atoms, and (iii) readout the final-state $\ket{\Psi (t)}$ after time $t$. We used the probability $P_0(t)$ of all atom back to the initial state, defined by
\begin{equation}
P_0(t)=|\bra{W_0} \Psi (t) \rangle|^2=|\bra{W_0}e^{-i\hat{H}t/\hbar} \ket{W_0}|^2
\end{equation}
where $\ket{W_0}$ is the initial zero-excitation state, e.g., $\ket{W_0}=\ket{000}$ for $N=3$, $\ket{0000}$ for $N=4$, and $\ket{000000}$ for $N=6$. With  eigenenergies $\lambda_j$, $P_0(t)$ is given by
\begin{equation}
\label{P0}
P_0(t) = \sum_{j} \left |A_j  \right |^4  + \sum_{j<k} B_{jk} \cos(\lambda_{jk}t),
\end{equation}
with $A_j = \bra{W_0}\lambda_j \rangle$ and $B_{jk}=2 \left|A_j \right|^2 \left|A_k \right|^2$. So the Fourier transform, $\mathcal{F}[P_0(t)]$, retrieves the energy differences, $\lambda_{jk}=\lambda_j - \lambda_k$ for all pairs of eigenenergies.  

Experiments were performed with an updated version of the machine previously used in our earlier works for deterministic multi-atom arrangements~\cite{lee2016, kim2016,lee2017}, quantum many-body thermalization~\cite{kim2018}, and entanglement generations~\cite{jo2020}. For the current work, we have increased the Rabi coherence time from 2.5~$\mu$s~\cite{lee2019} to about 10~$\mu$s, and developed three-dimensional atom arrangements, through technical improvements to be discussed in Sec. V.  The procedures of (i)-(iii) are summarized below:
 
(i) Atom arrangement: Single atoms are trapped with optical tweezers and arranged in three dimensional space. Rubidium ($^{87}$Rb) atoms first are cooled below 30~$\mu$K by Doppler and polarization gradient cooling in a magneto-optical trap (MOT), and optically pumped to the ground hyperfine state $\ket{0}=\ket{5S_{1/2}, F=2, m_F=2}$. Then, a spatial light modulator (SLM, Meadowlarks 512$\times$512 XY modulator) turns on as-many-as 250 optical tweezers (off-resonant optical dipole traps) to capture and rearrange $N$ single atoms deterministically to target positions, with 5-10~$\mu$m spacing~\cite{lee2017,kim2019}. The wavelength of the optical tweezers is $820~$nm and an objective lens (Mitutoyo G Plan Apo 50$\times$) of a high numerical aperture (NA $=0.5$) is used. The trap depth and diameter are 1~mK and 2~$\mu$m, respectively, and the lifetime of each trapped atom is about 40~s. 

(ii) Implementation of $H$: With the optical tweezers turned off, the Hamiltonian $H$ in Eq.~\eqref{H1} is implemented, through the Rydberg-atom two-photon excitation from $\ket{0}$ to $\ket{1}=\ket{71S_{1/2}, m_J=1/2}$ via the off-resonant intermediate state $\ket{m}=\ket{5P_{3/2}, F'=3, m_F'=3}$. For the two-photon excitation, we use 780-nm (Toptica DL Pro) and 480-nm (Toptica TA-SHG Pro) lasers, of which the beams counter-propagate with diameters of about $180~\mu$m and $35~\mu$m, respectively, and the laser frequencies are stabilized to a linewidth less than $(2\pi)30$~kHz, with a reference cavity (from Stable Laser Systems of finesse 15,000). The Rabi frequency of the given two-photon transition is given by $\Omega=\Omega_{0m}\Omega_{m1}/(2\Delta_{0m})$, when the intermediate detunning is $\Delta_{0m}=-(2\pi)600$~MHz. Measured Rabi frequencies, calibrated with reference single atoms, is $\Omega=(2\pi)1.0(1)$~MHz for $N=3, 4$ experiments  in Sec. IV and $(2\pi)0.8(1)$~MHz for the $N=6$ experiment. Correspondingly, the Rydberg blockade radii are given by $r_b=|C_6/\hbar \Omega|^{1/6}=10(1)$~$\mu$m and $11(1)$~$\mu$m, respectively.

(iii) Final state readout: In the detection stage, the optical tweezers are turned back on and the fluorescence from trapped ground-state atoms in $\ket{0}$ is collected through the same objective lens and imaged onto an EMCCD camera. The tomographic images of a 3D atomic array structure are obtained with an electrically focus-tunable lens (ETL, EL-16-40-TC from Optotune), located after the tube lens, by sequentially shifting the focal length. At the same time, the EMCCD is triggered on and off with a period of 40~ms (camera exposure time), for each tomogram. After the interaction $H$ of duration $t$, up to 5~$\mu$s with time step $\Delta t= 0.1$~$\mu$s, the events of all atoms back in the bare-atom ground states are collected, and the whole procedure, (i)-(iii), is repeated about 100-200 times of data accumulation to obtain the $\ket{W_0}$-state probability in Eq.~\eqref{P0}.

\section{Results and analysis}

In the first experiment, we probe three-atom configurations. As in Fig.~2(a), three atoms, $ABC$, are initially arranged in the triangle configuration, with $\overline{AB}=\overline{BC}=d=8$~$\mu$m, and the bending angle $\theta=\angle ABC$ is gradually changed from 60$^\circ$ (a triangle) to 180$^\circ$ (a linear chain). The corresponding atom positions are $A(-d,0,0)$, $B(0,0,0)$, and $C(-d\cos\theta, d\sin\theta,0)$ in Cartesian coordinates. The energy levels, given by the direct diagonalization of $H$, are shown in Fig.~2(b), in which the bright states ($\ket{\lambda_1}$, $\ket{\lambda_2}$, $\ket{\lambda_4}$, and $\ket{\lambda_5}$) are depicted with solid lines and the dark state ($\ket{\lambda_3}$, others are out of the given spectral range) with a dashed line. 

\begin{figure}[tbhp]
\centering
\includegraphics[width=0.45\textwidth]{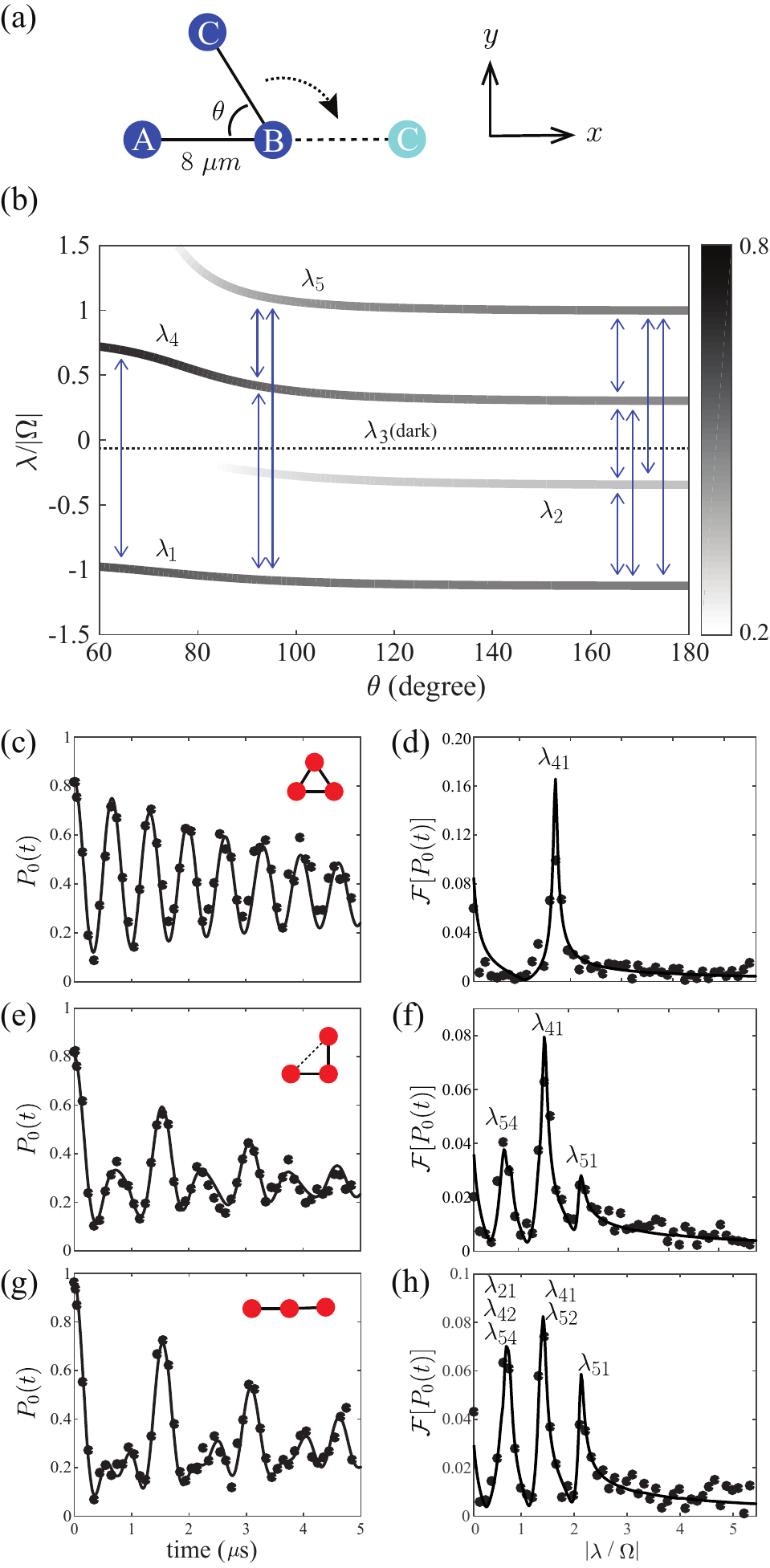}
\caption{(Color online) (a) Configuration of three atoms, changed from a triangle 
($\theta=60^\circ$) to a linear chain ($\theta=180^\circ$). (b) Enegy levels of ${H}$ in Eq.~\eqref{H1} applied for the given three-atom configurations. (c) Measured time-evolution of $P_0(t)=|\bra{000}\Psi(t)\rangle|^2$ for $\theta=60^\circ$, and (d) its Fourier transform. (e,f) $\theta=90^\circ$. (g,h) $\theta=180^\circ$. In (c-h), closed circles are experimental and lines are theoretical.}
\label{3atom}
\end{figure}

There are three characteristic regimes: (i) $U_{AC} > \Omega$, the super-atom regime near $\theta=60^\circ$, (ii) $U_{AC} \sim \Omega/10$, the $AC$ double-excitation regime near $\theta=90^\circ$, and (iii) $U_{AC} \ll \Omega$, the linear chain regime near $\theta=180^\circ$. In the super-atom regime, (i) $U_{AC} > \Omega$, maximally one atom is excited among $ABC$, due to the blockade effect. Figure 2(c) plots the time evolution of the probability, $P_{0}(t)=|\langle000 \ket{\Psi(t)}|^2$, measured at $\theta=60^\circ$, which shows the collective Rabi oscillation with frequency $\Omega_c=\sqrt{3}\Omega$. The two eigenstates, of respective energies $\lambda_1=-\Omega_c/2$ and $\lambda_4=\Omega_c/2$, are constructed with two symmetry base states, $\ket{W_0}=\ket{000}$ (the zero-excitation state) and $\ket{W_1}=(\ket{100}+\ket{010}+\ket{001})/\sqrt{3}$ (the super-atom state). The Fourier transform is given in Fig.~2(d), showing the resonance agreeing with $\lambda_4-\lambda_0=\Omega_c$ within the linewidth limited by the coherence. In the second and third regimes (ii) and (iii), the three-atom symmetry is broken. The single-excitation state $\ket{W_1}$ is energy-splitted to $\ket{W_1'}=(\ket{100}+\ket{001})/\sqrt{2}$ and $\ket{010}$, and the double-excitation state $\ket{101}$ appears. As a result, there are two additional states $\ket{\lambda_2}$ and $\ket{\lambda_5}$. The energy level calculation in Fig.~2(b) shows that the higher-energy state $\ket{\lambda_5}$, that is almost $\ket{101}$ at $\theta\sim80^\circ$, starts to be coupled before $\ket{\lambda_2}$, indicating that the $AC$ double-excitation state $\ket{101}$ is generated in the second regime (ii) $U_{AC} \sim \Omega/10$. In agreement, the measurements in the regime (ii), at $\theta=90^\circ$, are plotted in Figs. 2(e,f), showing three resonances, $\lambda_{54}$, $\lambda_{41}$, and $\lambda_{51}$. In the regime (iii), measurements for $\theta=180^\circ$ in Figs. 2(g,h) show only three resonances, while six ($_4C_2=6$) are expected; however, numerical calculation of the linear chain ($U_{AC} \ll \Omega$) confirms that the $\ket{\lambda_2}$ amplitude is nonzero and that two of the measured resonances are energy degenerate, $\lambda_{54}=\lambda_{42}=\lambda_{21}$ and $\lambda_{41}=\lambda_{52}$, within the spectral resolution. The lines in Figs.~2(c-h) are numerical simulations of the given three-atom dynamics, carried out with the method introduced in Refs.~\cite{lee2019,nakagawa2020}, in which Lindbladian master equations are solved with state-preparation-and-measurement errors, intrinsic dephasing, and laser spectral noises taken into account.

\begin{figure*}[tbp]
\centering
\includegraphics[width=0.95\textwidth]{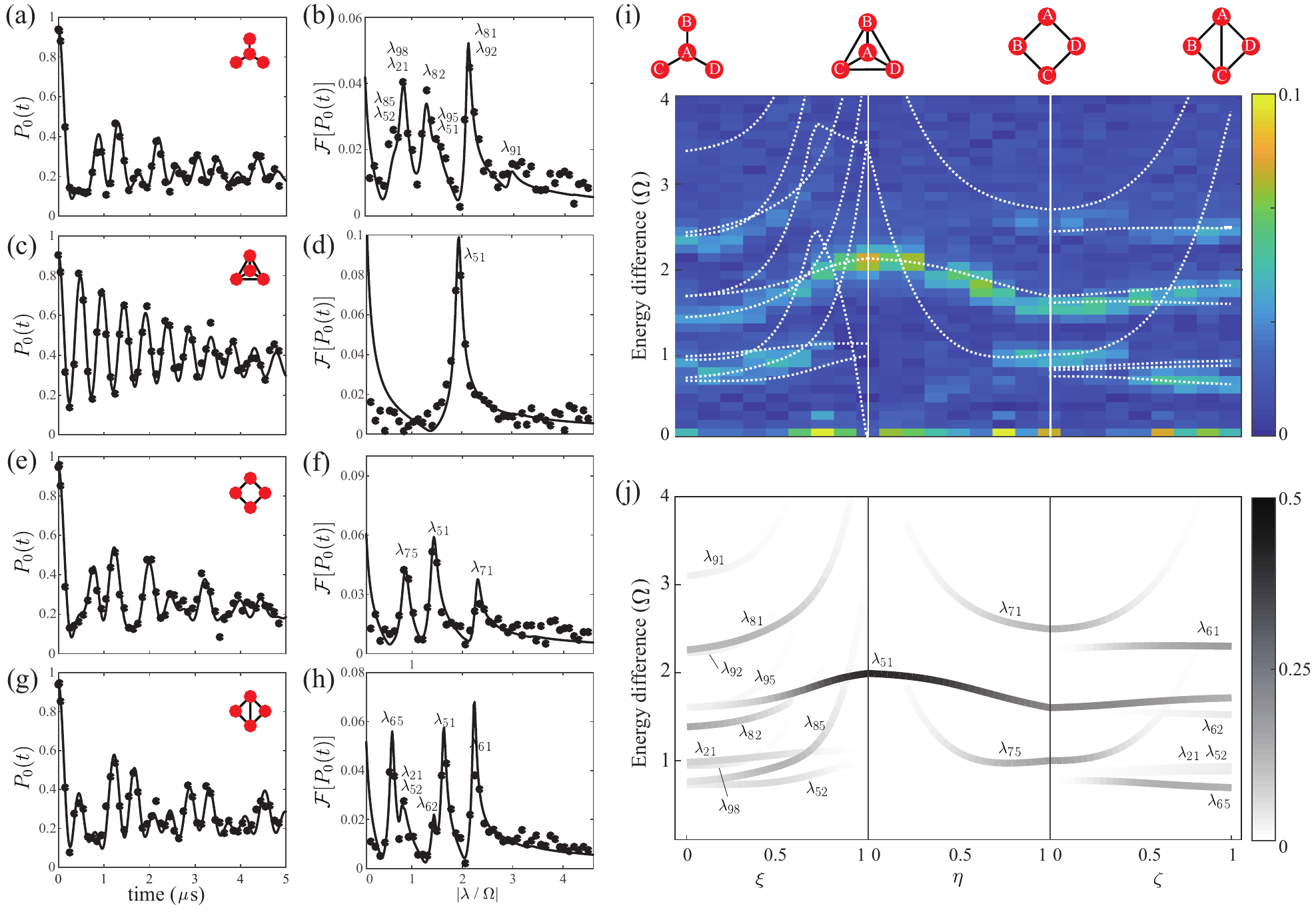}
\caption{(Color online) Structural transformation of an $N=4$ Rydberg-atom system in the sequence of $S_4 \rightarrow K_4 \rightarrow C_4 \rightarrow K_4$-e. (a,b) The star graph $S_4$: (a) Measured $P_0(t)=|\bra{0000}\Psi(t)\rangle|^2$; and (b) Fourier transform $\mathcal{F}[P_0(t)]$. (c,d) The complete graph $K_4$. (e,f) The cyclic graph $C_4$. (g,h) The diamond graph $K_4$-e. (i) Measured energy differences $\mathcal{F}[P_0(t)]$ vs. the deformation parameters ($\xi$, $\eta$, and $\zeta$). (j) Theoretical eigenenergy differences $\lambda_{jk}$ (without decoherence effects taken into account), in which the solid and dashed lines denote bright and dark transitions, respectively, and the gray scale represents the power spectral density.}
\label{four_example}
\end{figure*}

Now, in the second experiment, we probe the structural transformations of the $N=4$ atom system introduced in Fig.~1. Measured results for the four 4-vertex-connected graphs $S_4$, $K_4$, $C_4$, and $K_4$-e, are shown in Figs.~3(a,b), 3(c,d), 3(e,f), and 3(g,h), respectively. The time evolution of the $\ket{W_0}$-state probability, $P_0(t)=|\bra{0000}\Psi(t)\rangle|^2$,  and their Fourier transforms, $\mathcal{F}[P_0(t)]$, are shown (with closed circles) in comparison with numerical calculations (with lines). Mainly four spectral peaks are observed for $S_4$, which correspond to the nondegenerate energy differences, $\lambda_2-\lambda_1=\lambda_9-\lambda_8$, $\lambda_8-\lambda_2$, $\lambda_9-\lambda_2=\lambda_8-\lambda_1$, and $\lambda_9-\lambda_1$ in Table~I.  Likewise, $K_4$ has one peak, $\lambda_5-\lambda_1$, and $C_4$ has three, $\lambda_7-\lambda_5$, $\lambda_5-\lambda_1$, and $\lambda_7-\lambda_1$. $K_4$-e has three peaks $\lambda_6-\lambda_5$, $\lambda_5-\lambda_1$, and $\lambda_6-\lambda_1$. We note that additional peaks are identified by numerical calculations (lines) in Fig.~3(d) and 3(h) due to the next nearest neighbor couplings involved with $\lambda_5$ of $S_4$ and $\lambda_2$ of $K_4$-e, respectively. 

Structural transformations are performed through the sequence of $S_4 \rightarrow K_4 \rightarrow C_4 \rightarrow K_4$-e, as introduced in Fig.~1. The first transformation $S_4 \rightarrow K_4$ in Fig.~1(b) is from a star to a tetrahedron, which is pulling out the center atom of $S_4$ from the plane of the rest three atoms, while the lengths of the three edges are maintained, until six edges of an equal length are formed. Suppose initial atom positions are $(0,0,0)$ and $(\cos\theta_i,\sin\theta_i,0)$ with $\theta_{1,2,3}=0, 2\pi/3, 4\pi/3$. Then, the given transformation can be parameterized with new positions $(0,0,\sqrt{2/3}\xi)$ and $(1-(1-2/\sqrt{3})\xi)(\cos\theta_i,\sin\theta_i,0)$, where $\xi$ changes from 0 to 1. Similarly, the second transformation $K_4 \rightarrow C_4$ in Fig.~1(c) is from a tetrahedron to a square. It is stretching two non-adjacent edges of a tetrahedron, while keeping the lengths of the rest four edges the same, until a square is formed. This second structural change can be defined by the new locations $(-\eta/\sqrt{2}, 0, \sqrt{2/3}(1-\eta))$, $(\sqrt{1/3}+(1/\sqrt{2}-\sqrt{1/3})\eta, 0 ,0)$, $(-(1-\eta)/2\sqrt{3}, 1/2+(1/\sqrt{2}-1/2)\eta, 0)$, and $(-(1-\eta)/2\sqrt{3}, -1/2+(-1/\sqrt{2}+1/2)\eta, 0)$ parameterized with $\eta$ changing from 0 ($K_4$) to 1 ($C_4$). The last transformation $C_4 \rightarrow K_4$-e in Fig.~1(d) is deforming a square to a diamond, with parameterized locations of $(\mp 1/\sqrt{2} \mp (1/2-1/\sqrt{2})\zeta,0,0)$ and $(0,\pm 1/\sqrt{2} \pm (\sqrt{3}/2-1/\sqrt{2})\zeta,0)$ between $\zeta=0$ ($C_4$) and $\zeta=1$ ($K_4$-e). Experimental results are summarized in Fig.~4(i) for the given sequence of structural transformations. The Fourier transform $\mathcal{F}[P_0(t)]$ is plotted as a function of the parameters $\xi$, $\eta$, and $\zeta$:  $\xi=0$ corresponds to $S_4$ in Fig.~3(b), $\eta=0$ ($\xi=1$) to $K_4$ in Fig.~3(d), $\zeta=0$ ($\eta=1$) to $C_4$ in Fig.~3(f), and $\zeta=1$ to $K_4$-e in Fig.~3(h). The measured spectrum is compared to the numerical calculation with $H$, the superposed dashed lines in Fig.~3(i) and the gray-scale lines in Fig.~3(j). Within the Fourier transform resolution, the retrieved energies are in qualitative agreement with the theory.

\begin{figure*}[tbp]
\centering
\includegraphics[width=0.95\textwidth]{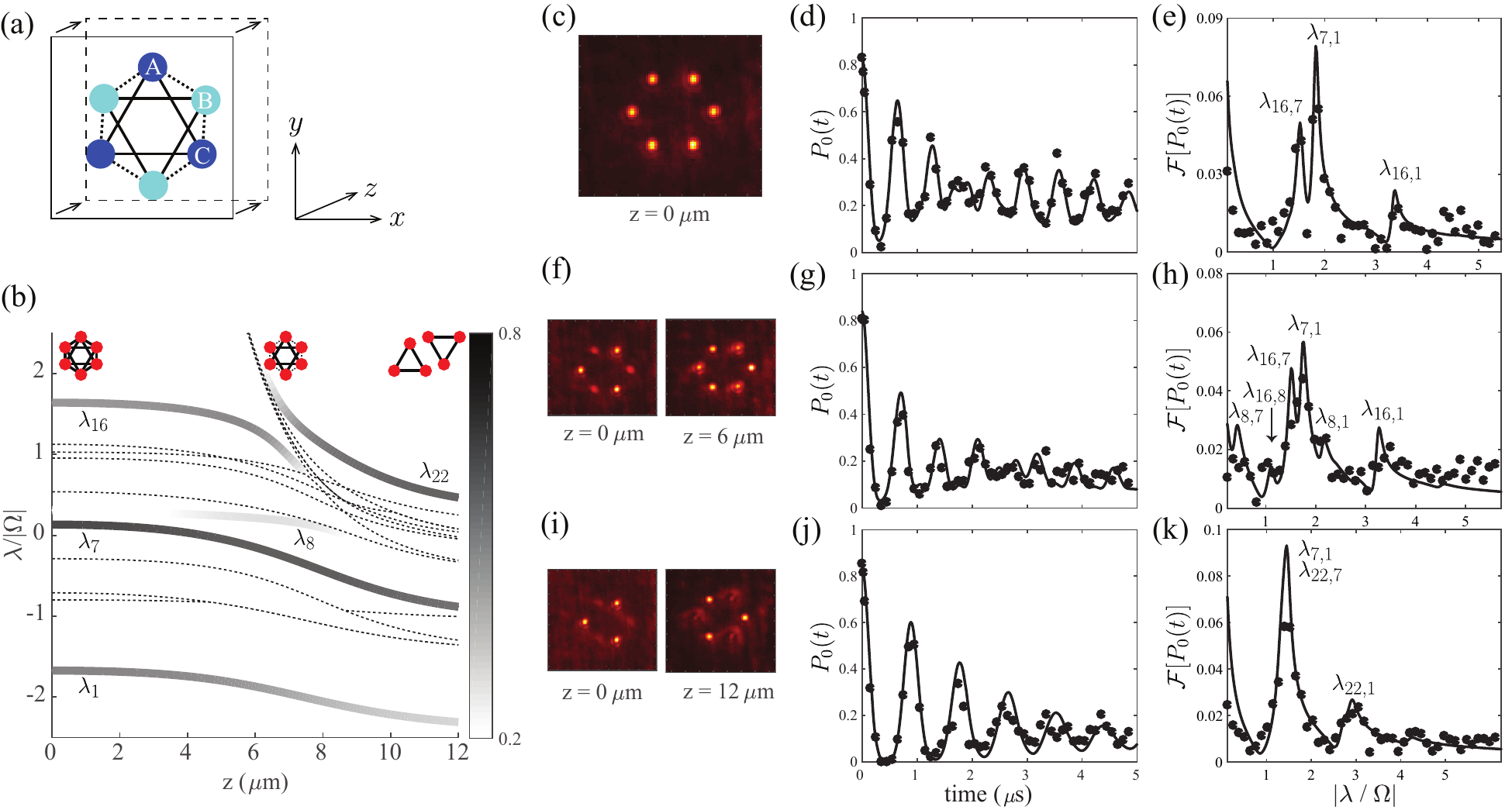}
\caption{(Color online) Spectroscopy of $N=6$ atoms: (a) An $N=6$ atom system is deformed from a hexagon to an antiprism, by axially separating two planar triangles. (b) Energy levels vs. plane separation ($z$). (c,d,e) The hexagon configuration at $z=0$: (c) Fluorescence image, (d) $P_0(t)=|\bra{0000000}\Psi(t)\rangle|^2$, (e) $\mathcal{F}[P_0(t)]$. (f,g,h) An antiprism with $z=3d/2$. (i,j,k) An antiprism with $z=3d/2$.}
\label{6atom}
\end{figure*}

In the final experiment, we probe the structural transformation of an $N=6$ atom system, from a hexagon to an antiprism (a set of  upright and inverted triangles, separated by $z$). As illustrated in Fig.~4(a), six atoms are initially arranged at the vertices of a hexagon, with positions $d(\cos\theta_i, \sin\theta_i,0)$ with $\theta_{j}=j\pi/3$ (for $j=1,\cdots,6$) and the axial $z$ positions of the even numbered atoms ($j=2,4,6$) are axially translated from $z=0$ to $z=3d/2$. During the transformation, the length of each triangle is kept constant, $\overline{AC}=d=8$~$\mu$m, and the distance $\overline{AB}$ is changed as $\overline{AB}(z)=\sqrt{d^2/3+z^2}$. The energy levels calculated with $H$ are plotted in Fig.~4(b), which shows three distinct coupling regimes: (i) $U_{AB} 
\gg \Omega$, the hexagon regime at $z=0$, (ii) $U_{AB} \sim \Omega$, the $AB$ double-excitation regime around $z=3d/4$, and (iii) $U_{AB} \ll \Omega$, the decoupled trios near $z=3d/2$.

In the hexagon regime, (i) $U_{AB} \gg \Omega$, there are three eigenstates, constructed with symmetric base states $\ket{W_0}=\ket{000000}$,  $\ket{W_1}$(the superposition of single-excitation states), and $\ket{W_2^d}=(\ket{100100}+\ket{010010}+\ket{001001})/\sqrt{3}$ (the superposition of diagonal double excitations). We denote the eigenstates by $\ket{\lambda_1}$, $\ket{\lambda_7}$, and $\ket{\lambda_{16}}$. The results for the six atoms in the hexagon configuration at $z=0$ are shown in Figs.~4(c,d,e) with the fluorescence image, the measured $\ket{W_0}$ probability, and the Fourier transform, respectively. Likewise, the results for $z=3d/4$ and $z=3d/2$ are given in Figs.~4(f-h) and 4(i-k), respectively. At $z=3d/4$, which we refer to as (ii) the $AB$ double-excitation regime ($U_{AB} \sim \Omega$), the distance between $AB$ atoms is bigger than the blockade radius, i.e., $\overline{AB}>r_b$, so $AB$ can be excited together, while they are weakly coupled ($U_{AB} \sim \Omega$). Therefore, besides the above base states, $\ket{W_0}$, $\ket{W_1}$, and $\ket{W_2^d}$, an additional symmetric base state, $\ket{W_2^{AB}}$ (the superposition of adjacent double excitations), is allowed. The spectrum in Fig.~4(h) shows $_4C_2=6$ peaks, in a reasonable agreement with the numerical calculation. In the regime (iii)  $U_{AB} \ll \Omega$, at $z=3d/2$, the atom planes are well separated, and, as a result, the two sets of three atoms are decoupled, each constructed with its own symmetric basis, $\ket{000}$ and $(\ket{100}+\ket{010}+\ket{001})/\sqrt{3}$. So the eigenenergies, of the decoupled trios, are given by $\lambda_1=-\sqrt{3}\Omega$, $\lambda_7=0$, and $\lambda_{22}=\sqrt{3}\Omega$.

\section{Technical details and improvements}

The spectral resolution of the given spectroscopy is limited by the coherence time of the Rydberg-atom quantum simulator. It is discussed in Refs.~\cite{lee2019,nakagawa2020} that the coherent operation time of the machine is dominantly limited by the non-intrinsic dephasing due to laser spectral phase noises. In order to suppress the laser spectral phase noises, we adopted the laser frequency stabilization method described in Ref.~\cite{gouet2009} without using intra-cavity electro-optic modulations in this work. The laser frequency was locked to a resonance of a high-finesse cavity using Pound-Drever-Hall (PDH) technique. The reflected light from the cavity was directed into a PDH module (Stable Laser System PDH-1000-20D) which included a fast photo-detector and electronics to demodulate the detected beat signal. The demodulated signal was fed into a fast analog proportional-integral-derivative (PID) controller (Toptica FALC 110) and two separate (`slow' and `fast') servo loops were implemented; `slow' for changing the angle of the grating in the extra-cavity diode laser, mostly to compensate frequency drift, and `fast' for changing the current through the laser diode, mostly to reduce the linewidth. By setting the PID parameters (to get the highest possible gain for low frequency and relatively low gain for high frequency with a proper amount of $90^{\circ}$ phase shifted signals) and optimizing the transfer function of the servo loop, we restrained the oscillation of the servo-loop at the margin of the bandwidth (which otherwise caused a servo-bump) and achieved the sufficiently low frequency noise level, $S_{\nu}(f)<10^3$ Hz$^2$/Hz, even at a servo-bump around 1 MHz. With the described method, the coherence time measured from single-atom Rabi oscillation decay is improved from 2.5~$\mu$s~\cite{lee2019} to 10~$\mu$s.

To make the atomic arrangements of the given geometries, we extended the method of dynamic holographic optical tweezers~\cite{kim2019}, previously restricted to 2D arrangements, to a 3D version. The hologram on demand for an optical tweezer arrangement was calculated with a 3D Gerchberg-Saxton (GS) algorithm, along with the methods of weighted-GS and phase induction~\cite{kim2019} for fast convergence.  For each cycle of atom rearrangement, positions of about 26 optical tweezers were simultaneously shifted by differential displacements frame-by-frame, throughout a serial sequence of 35 successive phase patterns in 700~ms, to achieve the occupation probability per site $>0.94$. While the transverse fluctuation of trap positions was below the imaging resolution limit, the axial position fluctuation was about 1~$\mu$m, due to limited phase convergence. The number of GS-algorithm iterations was set to five in experiments, compromising between the quality of the optical tweezers and the calculation time. 

A typical time budget for an experiment with a single-plane arrangement of atoms is less than  one second, given by the sum of the times for atom loading (100~ms), initial occupancy checking (40~ms), atom rearrangements (700~ms), final occupancy checking (40~ms), optical pumping (2~ms), Rydberg-atom excitation (5~$\mu$s), and final state detection (40~ms). When atom arrangements are repeated for a multi-plane geometry, the overal time increases but is little significant compared to the 40-s trap life time.

\section{Conclusion}
In summary, we have utilized three-dimensional arrangements of neutral atoms, of adjustable inter-atom distances, to obtain the conformation energy landscape of strongly-interacting, small-scale Rydberg atom systems, in particular, during their structural transformations. We probed all possible nonisomorphic, connected graph configurations programmed for $N=3,4$ atoms, and partial graphs for $N=6$. The experimentally measured topology-dependent eigenspecta are in good agreement with the model calculation of the few-body quantum-Ising Hamiltonian. It is hoped that high-dimensional programming of qubit connectivities demonstrated in this paper shall be useful for  further applications of programmable quantum simulators.

\begin{acknowledgements}
This research was supported by Samsung Science and Technology Foundation (SSTF-BA1301-52), National Research Foundation of Korea (NRF) (2017R1E1A1A01074307), and Institute for Information \& Communications Technology Promotion (IITP-2018-2018-0-01402).
We thank Woojun Lee, Hansub Hwang, and Heekun Nho for assistance in constructing 3D optical tweezer traps.
\end{acknowledgements}


\begin{thebibliography}{1}


\bibitem{Feynman1982} R. Feynman, ``Simulating physics with computers,'' Int. J. Theor. Phys.~{\bf 21}, 467 (1982).

\bibitem{Nielsen2000} M. A. Nielsen and I. K. Chuang, \textit{Quantum Information and Quantum Computation} (Cambridge University Press, Cambridge, 2000).

\bibitem{Georgescu2014} I. M. Georgescu, S. Ashhab, and F. Nori, ``Quantum simulation,'' Rev. Mod. Phys.~\textbf{86}, 153 (2014).

\bibitem{Gross2017} C. Gross and I. Bloch, ``Quantum simulations with ultracold atoms in optical lattices,'' Science {\bf 357}, 995 (2017).


\bibitem{Duan2010} L.-M. Duan and C. Monroe, ``\textit{Colloquium:} Quantum networks with trapped ions,'' Rev. Mod. Phys.~\textbf{82}, 1209 (2010).

\bibitem{Wendin2017} G. Wendin, ``Quantum information processing with superconducting circuits: a review,'' Rep. Prog. Phys.~\textbf{80}, 10 (2017).

\bibitem{briegel2000} H. J. Briegel, T. Calarco, D. Jaksch, J. I. Cirac, and P. Zoller, ``Quantum computing with neutral atoms,'' J. Mod. Opt. \textbf{47}, 415 (2000).

\bibitem{Weiss2017} D. S. Weiss and M. Saffman, ``Quantum computing with neutral atoms,'' Phys. Today \textbf{70}, 44 (2017).


\bibitem{Saffman2016} M. Saffman, ``Quantum computing with atomic qubits and Rydberg interactions: progress and challenges,'' J. Phys. B: At., Mol. Opt. Phys.  {\bf 49}, 202001 (2016).


\bibitem{Lukin2001} M. D. Lukin, M. Fleischhauer, R. Cote, L. M. Duan, D. Jaksch, J. I. Cirac, and P. Zoller, ``Dipole blockade and quantum information processing in mesoscopic atomic ensembles,'' Phys. Rev. Lett. {\bf 87}, 037901 (2001).

\bibitem{wilk2010} T. Wilk, A. Ga\"{e}tan, C. Evellin, J. Wolters, Y. Miroshnychenko, P. Grangier, and A. Browaeys, ``Entanglement of two individual neutral atoms using Rydberg blockade,'' Phys. Rev. Lett. {\bf 104}, 010502 (2010).

\bibitem{isenhower2010} L. Isenhower, E. Urban, X. L. Zhang, A. T. Gill, T. Henage, T. A. Johnson, T. G. Walker, and M. Saffman, ``Demonstration of a neutral atom controlled-NOT quantum gate,'' Phys. Rev. Lett. \textbf{104}, 010503 (2010).


\bibitem{Omran2019} A. Omran, H. Levine, A. Keesling, G. Semeghini, T. T. Wang, S. Ebadi, H. Bernien, A. S. Zibrov, H. Pichler, S. Choi, J. Cui, M. Rossignolo, P. Rembold, S. Montangero, T. Calarco, M. Endres, M. Greiner, V. Vuleti\'{c}, and M.D. Lukin, “Generation and manipulation of Schr\"{o}dinger cat states in Rydberg atom arrays,” Science, {\bf 365}, 570 (2019).


\bibitem{Marcuzzi2017} M. Marcuzzi, J. Min\'{a}r, D. Barredo, S. de L\'{e}s\'{e}leuc, H. Labuhn, T. Lahaye, A. Browaeys, E. Levi, and I. Lesanovsky, ``Facilitation dynamics and localization phenomena in Rydberg lattice gases with position disorder,'' Phys. Rev. Lett. {\bf 118}, 063606 (2017).


\bibitem{kim2018} H. Kim, Y. J. Park, K. Kim, H. S. Sim, and J. Ahn, ``Detailed balance of thermalization dynamics in Rydberg-atom quantum simulators,''
Phys. Rev. Lett. {\bf 120}, 180502 (2018).


\bibitem{Bernien2016} H. Bernien, S. Schwartz, A. Keesling, H. Levine, A. Omran, H. Pichler, S. Choi, A. S. Zibrov, M. Endres, M. Greiner, V. Vuleti\'{c}, and M. D. Lukin, ``Probing many-body dynamics on a 51-atom quantum simulator,'' Nature {\bf 551}, 579 (2016).

\bibitem{Sylvain2019} S. de L\'{e}s\'{e}leuc, V. Lienhard, P. Scholl, D. Barredo, S. Weber, N. Lang, H.P. B\"{u}chler, T. Lahaye, and A. Browaeys, ``Observation of a symmetry protected topological phase of interacting bosons with Rydberg atoms,'' Science {\bf 365}, 775 (2019).





\bibitem{Browaeys2020} A. Browaeys and T. Lahaye, ``Many-body physics with individually controlled Rydberg atoms,'' Nat. Phys. {\bf 16}, 132 (2020).


\bibitem{kim2016}  H. Kim, W. Lee, H.-g. Lee, H. Jo, Y. Song, and J. Ahn, ``In situ single-atom array synthesis by dynamic holographic optical tweezers,'' Nat. Commun. {\bf 7}, 13317 (2016).

\bibitem{barredo2016} D. Barredo, S. de L\'{e}s\'{e}leuc, V. Lienhard, T. Lahaye, and A. Browaeys, ``An atom-by-atom assembler of defect-free arbitrary 2d atomic arrays,'' Science {\bf 354}, 1021  (2016).

\bibitem{endres2016} M. Endres, H. Bernien, A. Keesling, H. Levine, E. R. Anschuetz, A. Krajenbrink, and M. D. Lukin, ``Atom-by-atom assembly of defect-free one-dimensional cold atom arrays,'' Science \textbf{354},  1024 (2016).

\bibitem{Arute2019} F. Arute et. al., ``Quantum supremacy using a programmable superconducting processor,'' Nature {\bf 574}, 505 (2019).

\bibitem{lee2016} W. Lee, H. Kim, and J. Ahn, ``Three-dimensional rearrangement of single atoms using actively controlled optical microtraps,'' Opt. Express \textbf{24}, 9816 (2016).

\bibitem{lee2017} W. Lee, H. Kim, and J. Ahn, ``Defect-free atomic array formation using the Hungarian matching algorithm,'' Phys. Rev. A {\bf 95}, 053424 (2017).

\bibitem{jo2020} H. Jo, Y. Song, M. Kim, and J. Ahn, ``Rydberg atom entanglements in the weak coupling regime,'' Phys. Rev. Lett. {\bf 124}, 033603 (2020).

\bibitem{lee2019} W. Lee, M. Kim, H. Jo, Y. Song, and J. Ahn, "Coherent and dissipative dynamics of entangled few-body systems of Rydberg atoms," Phys. Rev. A {\bf 99}, 043404 (2019).

\bibitem{nakagawa2020} H. Tamura, T. Yamakoshi, and K. Nakagawa, ``Analysis of coherent dynamics of a Rydberg-atom quantum simulator,''
Phys. Rev. A {\bf 101}, 043421 (2020).

\bibitem{kim2019} H. Kim, M. Kim, W. Lee, and J. Ahn, ``Gerchberg-Saxton algorithm for tweezer-trap atom arrangements,'' Opt. Express {\bf 27}, 2184 (2019).



\bibitem{gouet2009} J. Le Gou\"{e}t, J. Kim, C. Bourassin-Bouchet, M. Lours, A. Landragin, and F. Pereira Dos Santos, ``Wide bandwidth phase-locked diode laser with an intra-cavity electro-optic modulator,'' Opt. Commun. {\bf 282}, 977 (2009).

\end{thebibliography}
\end{document}